\documentclass[12pt,preprint]{aastex}
\usepackage{rotating,epsfig,graphicx}

\usepackage[left=2.5cm,top=2.55cm,right=2.55cm,bottom=2.45cm]{geometry}

\newcommand{\nsfsect}{\vspace{-0.5cm} \section}
\newcommand{\nsfsubsect}{\vspace{-0.5cm} \subsection}

\def\deg{\ifmmode^\circ\else$^\circ$\fi}  
\def\arcsec{\ifmmode {'' }\else $'' $\fi}  
\def\arcmin{\ifmmode {' }\else $' $\fi}    
\def\Msun       {${\rm M}_\odot$}
\def\ga{\mathrel{\hbox{\rlap{\hbox{\lower4pt\hbox{$\sim$}}}\hbox{$>$}}}}
\def\la{\mathrel{\hbox{\rlap{\hbox{\lower4pt\hbox{$\sim$}}}\hbox{$<$}}}}
\def\HI{{\mbox{\sc H i}}}

\begin{document}

\title{How do Galaxies Accrete Gas and Form Stars?}

\author{M.E. Putman (Columbia), P. Henning (UNM), A. Bolatto (U. Maryland), D. Keres (Harvard), D.J. Pisano (WVU/NRAO), J. Rosenberg (George Mason U.), F. Bigiel (UC-Berkeley), G. Bryan (Columbia), D. Calzetti (U. Mass), C. Carilli (NRAO), J. Charlton (PSU), H.-W. Chen (U. Chicago), J. Darling (U. Colorado), S. Gibson (W. Kentucky), N. Gnedin (FNAL), O. Gnedin (U. Michigan), F. Heitsch (U. Michigan), D. Hunter (Lowell), S. Kannappan (UNC), M. Krumholz (UC-Santa Cruz), A. Lazarian (U. Wisconsin), J. Lazio (NRL), A. Leroy (MPIA), F.J. Lockman (NRAO), M. Mac Low (AMNH), A. Maller (CUNY), G. Meurer (JHU), K. O'Neil (NRAO), J. Ostriker (Princeton), J.E.G. Peek (UCB), J.X. Prochaska (UC-Santa Cruz), R. Rand (UNM), B. Robertson (U. Chicago), D. Schiminovich (Columbia), J. Simon (Carnegie), S. Stanimirovic (UW-Madison), D. Thilker (JHU), C. Thom (U. Chicago), J. Tinker (UCB), J.M. van der Hulst (Groningen), B. Wakker (UW-Madison), B. Weiner (Arizona), A. Wolfe (UCSD), O.I. Wong (Yale), L. Young (NMT)											} 

\hfill
\vfill
\clearpage

Great strides have been made in the last two decades in determining how galaxies
evolve from their initial dark matter seeds to the complex structures we observe at $z=0$.
The role of mergers has been documented through both observations and simulations, numerous satellites that may represent these initial dark matter seeds have been discovered in the Local Group, high redshift galaxies have been revealed with monstrous star formation rates, and the gaseous cosmic web has been mapped through absorption line experiments.  Despite these efforts, the dark matter simulations that include baryons are still unable to accurately reproduce galaxies.  One of the major problems is our incomplete understanding of how a galaxy accretes its baryons and subsequently forms stars.  Galaxy formation simulations have been unable to accurately represent the required gas physics
on cosmological timescales, and observations have only just begun to detect the
star formation fuel over a range of redshifts and environments.
How galaxies obtain gas and subsequently form stars is a major unsolved, yet tractable problem in contemporary extragalactic astrophysics.  In this paper we outline how progress can be made in this area in the next decade.

\nsfsect{The Build-up of Baryons}
\label{high-z}

While the dark matter component of galaxies grows through mergers of
dark matter halos, the accretion of baryonic gas is more complex. In
addition to major and minor mergers, smoothly distributed gas - that
never collapsed into halos or was stripped from other galaxies - can
accrete onto galaxies directly from the intergalactic medium.
Depending on a halo's mass and redshift, gas can either shock-heat and form a
diffuse hot halo component, or flow deeply into the halo in the form of cold 
intergalactic filaments. Both direct accretion from the cold filaments
and cooling of the hot component can provide a gas supply for galaxies to form stars.
Hydrodynamical simulations including mergers and smooth accretion processes
are complex, but they are 
beginning to converge on the general features of gas accretion$^{1,2}$
and provide predictions for the relative importance of the various modes$^{3,4}$.
 
Results from recent simulations have already changed our understanding of
high redshift gas accretion with the discovery of cold mode accretion along intergalactic filaments (Figure 1, left). 
At lower redshift cooling of the gas in hot virialized halos, as
envisioned in classical models$^5$, starts to be an important source of
gas for galaxies. Currently such cooling is thought to proceed
through instabilities in the hot halo gas, leading to the formation of
cold gaseous clouds that fall towards the center of a halo and
provide a fresh gas supply to the ISM$^{6,7,8,9}$.  High resolution
simulations are beginning to demonstrate the formation of these clouds 
in a fully cosmological environment (Figure 1, right) and significant progress in this area
is expected in the next few years.  Direct detection of clouds will significantly constrain the models and
improve our understanding of low redshift gas accretion. 


\begin{figure}[h]
\hskip -12pt
\begin{center}
\epsfxsize=2.8in
\epsfbox{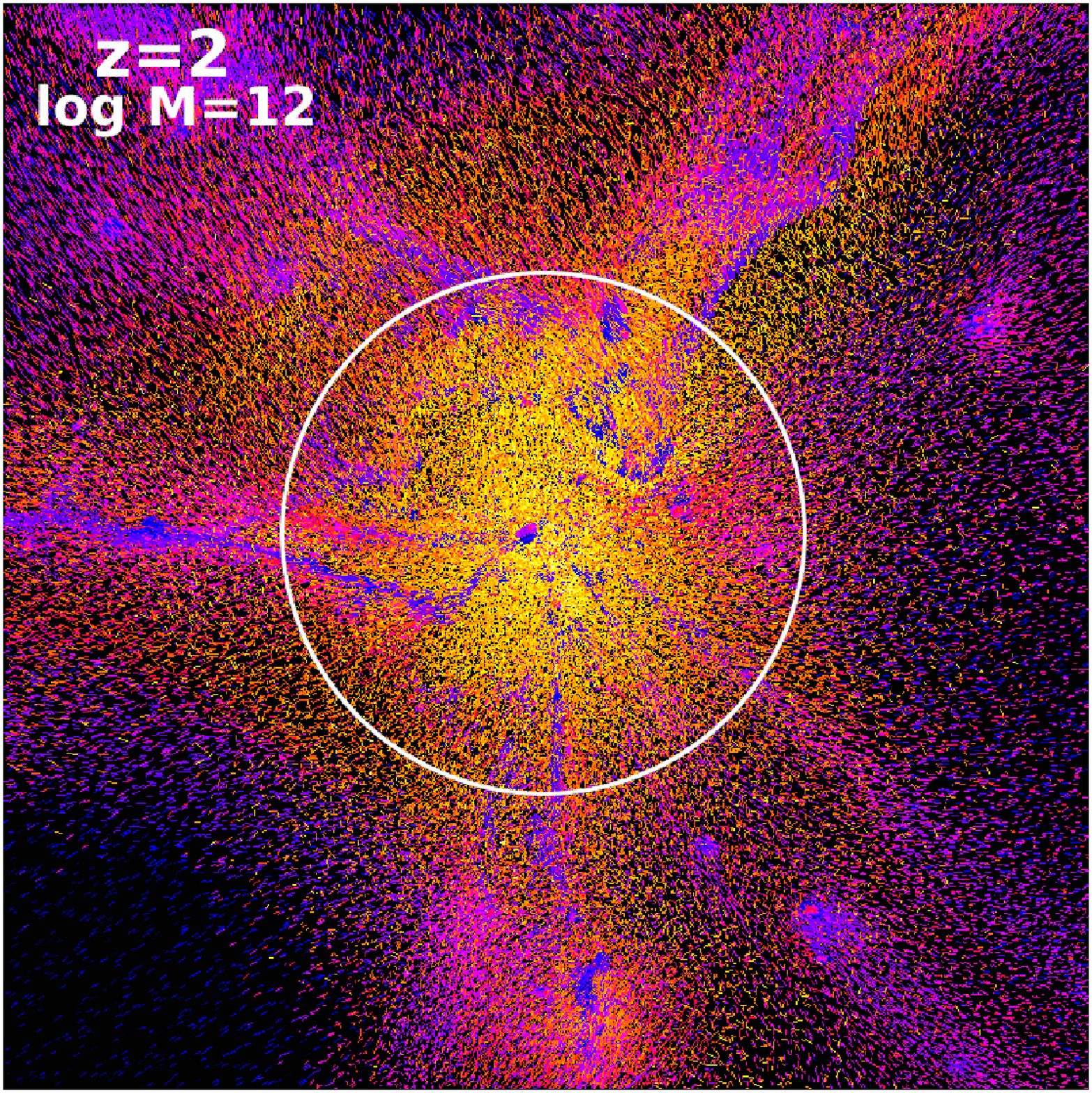}
\epsfxsize=2.8in
\epsfbox{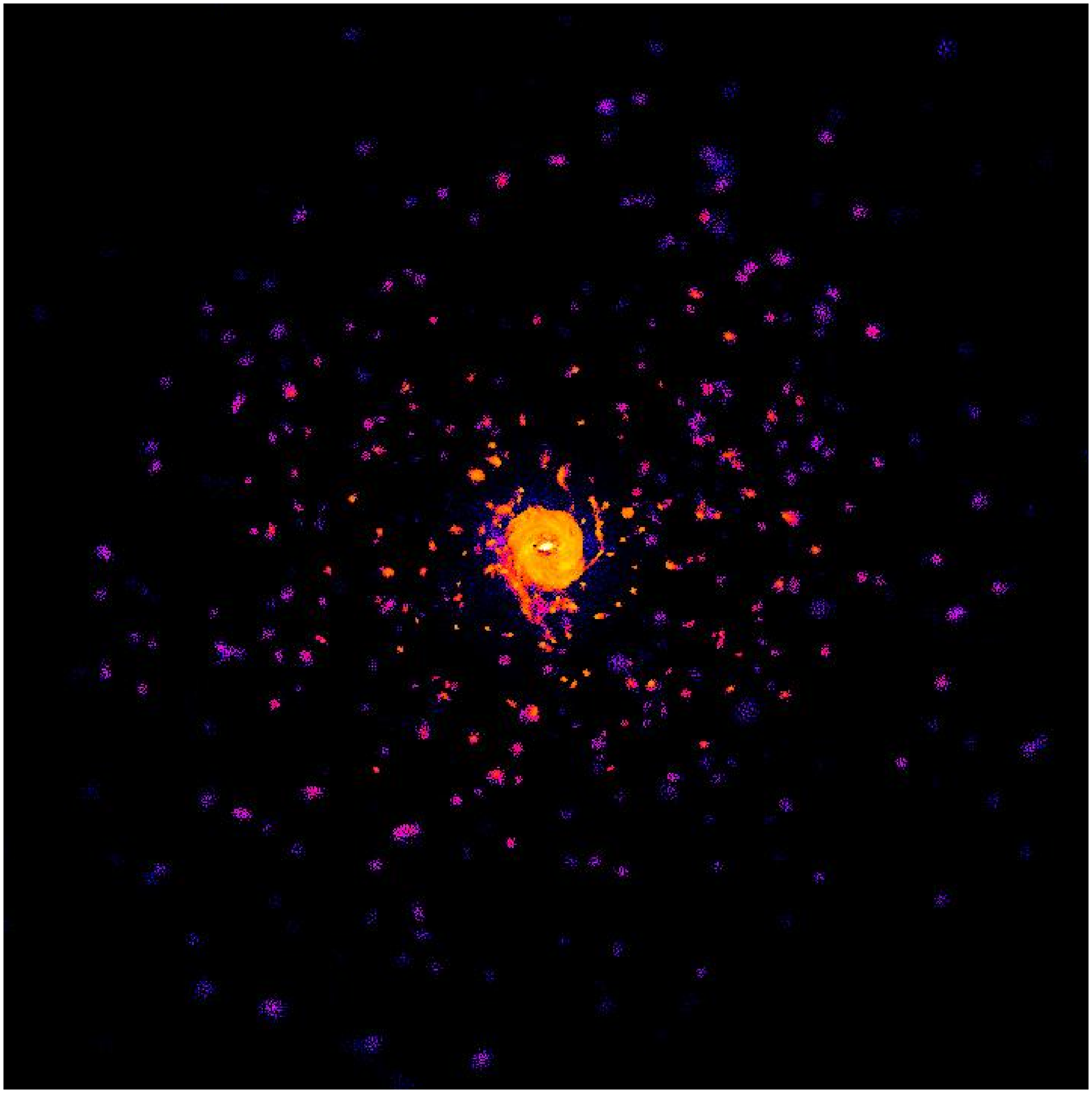}
\caption[]{\footnotesize {\it Left:} A Milky Way-size halo at $z=2$ being fed by 
 cold filamentary streams of gas. The temperature of the gas particles is color coded from
 blue ($\sim 10^4 \rm K$) to yellow ($\sim 10^6 \rm K$)$^2$. The circle indicates the virial radius
 of this halo. {\it Right:} A Milky Way-size halo at
 $z=0.1$. Only the dense gas is plotted revealing a wealth of cloud
 like objects. This region is 270 kpc on a side (from D. Kere\v{s}).}
 \end{center}
\end{figure}

\nsfsubsect{Observational Signatures of Gas Accretion}

Observations are beginning to reveal galaxies actively accreting star formation fuel, possibly with a hint of similarity to Figure 1 (right).   The Milky Way
has been known for some time to have halo clouds whose fate will be to fuel the disk$^{10,11}$.  
These clouds are thought to explain the metallicity of our Galaxy's stars$^{12}$ and its ability to continue forming stars at an average rate of 1-3 \Msun/yr.
Analogs to the Milky Way halo clouds have now been observed around other nearby spiral galaxies
through extremely deep \HI\ observations$^{13,14}$. These extragalactic analogs emphasize that gas accretion is ongoing at $z=0$, but
the origin of the accretion remains uncertain.   Some of the gas clouds may be satellite material$^{15,16}$,
while others may originate from the galaxy itself$^{17}$, or represent condensing density enhancements in the surrounding hot halo medium$^{18}$.  In addition, \HI\ clouds have been found in the vicinity of early-type galaxies with similar unknown origins$^{19,20}$.  
Given the hot gas reservoirs found around galaxies, the question remains if gas accretion ever actually stops (i.e., continual cooling onto the galaxy), or alternatively if the cold \HI\ clouds survive the trip to the galaxy through this medium$^{21,22}$.


Though observations capable of detecting the diffuse accreting gas are still limited, they have made it clear
that galaxy halos, the transition region between the cosmic filaments and 
the star forming disk, contain a complex, multi-phase medium in need of further study$^{23,24,25}$. 
 The origin of the accreting gas, the magnitude
of the accretion process, and how it may evolve with galaxy mass and redshift, remain key unknowns in a galaxy's build-up of baryons. 
 Future efforts will require multi-wavelength studies, including: deeper \HI\ observations that connect to the lower column density gas probed in absorption and are capable of detecting and resolving clouds at larger distances, observations that examine the dust content and metallicity of the gas, and kinematic studies of both the neutral and ionized gas to examine the transitions between the IGM, the halo, and the star forming galaxy.


\nsfsect{The Conversion of Gas to Stars}
\label{starform}

Once gas has accreted onto a galaxy we must understand how it
is then converted to stars to obtain a coherent picture of galaxy evolution.
Our weak grasp of the underlying physics of this process is
evident from some of the problems that arise in $\Lambda$CDM  galaxy formation simulations.
For instance, the excess of low-mass dark matter halos predicted compared to the number of low-mass galaxies observed
(the ``missing satellite problem''$^{26,27}$)
requires some type of suppression of star formation relative to Milky Way-size halos.
In addition,  it is unclear why the peak of the star formation activity is occurring in progressively smaller galaxies with time (i.e.,  ``downsizing''$^{28,29}$).  Several other key aspects
of the evolution of the universe are dependent on the
relation between gas and star formation. To name but a few: the existence of the blue sequence of
galaxies$^{30,31}$, the
progressive decline of the comoving SFR density in the universe since
$z\sim1$$^{32}$, the constancy of $\Omega_{HI}$ over time$^{33}$, the early enrichment
of the intergalactic medium$^{34}$, the mass-metallicity relation$^{35}$,
the low level of star formation associated with damped Ly-$\alpha$
systems$^{36}$, the formation of the extended UV disks$^{37}$,
and the nature of the initial mass function of stars$^{38}$.

\nsfsubsect{Star Formation in the Nearby Universe}

Studies of local galaxies are essential for determining the physical
processes responsible for converting the accreted atomic gas to molecular
gas and subsequently stars.  The
Milky Way appears to have enough gas to make another 5 billion suns, but we lack a
comprehensive picture of the processes that collect the low
density \HI\ into dense star-forming clouds.  The most popular
parametrization of the conversion of gas into stars is the
``Schmidt-law''$^{39,40}$, but the original explanation for
this relation in terms of large-scale
gravitational instabilities$^{41}$ faces a number of
challenges. The physical underpinning of the relation between gas
and star formation remains a matter of intense theoretical and observational research$^{42,43,44,45}$.

As an example, just in the last decade theories consistent with existing data
have argued for an array of physical processes regulating
star formation on a variety of spatial scales. Some of them 
address galactic scales to explain the collection of gas into molecular clouds$^{46,47,48,49,50}$, 
 others emphasize 
local processes regulating the star formation efficiency$^{51,52,53}$.
The numerous possible solutions accentuates the need for more sophisticated
modeling incorporating a wide range of physical processes,
and high resolution observations of a variety of star formation
and density regimes.

Observationally the goal is to accurately trace the star formation process from the initial
atomic hydrogen to the final stars.  
Recent observational progress in the field has come from employing UV,
optical, and FIR imaging to measure star formation rates$^{54,55,56,57}$,
and FIR, submm-, mm-, and cm-wave interferometric and
single-dish observations to characterize the warm and cold gas content
of galaxies$^{58,59,60,61}$. 
To date these studies are limited to just a handful of the brightest and nearest
galaxies (D$\lesssim$10 Mpc).  Obtaining high resolution, multi-wavelength observations 
in outer disks, low mass galaxies, and in more distant systems is required to determine the dominant processes dictating star formation.

\nsfsubsect{Census of Cold Gas and Stars Through Time}

A census of the cold gas and stellar components of galaxies over cosmic
time is another crucial component of understanding
how galaxies obtain gas and subsequently convert it to stars.
Figure 2 shows the relationship between the stellar, \HI, and star-formation rate
density as a function of redshift. The cosmological mass density of
\HI\ is nearly constant over the past $\sim$10 Gyr while the stellar density
continues to increase.   This shows the importance of ongoing gas accretion 
and the conversion of atomic to molecular gas
in the star formation process. Surveys of damped Ly-$\alpha$ systems (the points above $z=0.24$ in Figure 2) do not reveal how the gas is
distributed in relation to the ongoing and previous star formation in galaxies.  By connecting \HI\ gas
to the stars and star-formation rates in individual galaxies, one can constrain
the sequence of gas accretion and consumption.

\begin{figure}[th]
\begin{center}
\begin{minipage}[h]{0.7\linewidth}
\psfig{figure=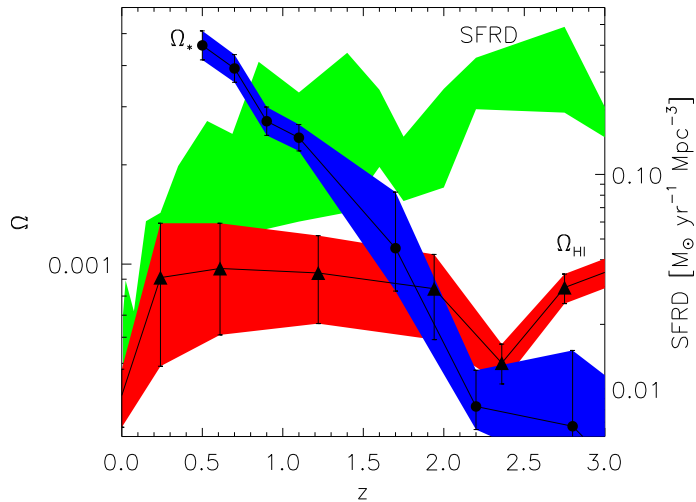,height=4.0in,angle=90}
\end{minipage}\begin{minipage}[h]{0.3\linewidth}
\caption{\footnotesize The relationship between the mass density in stars, $\Omega_{\rm stars}$ (blue$^{78,79}$), and in gas, $\Omega_{\rm HI}$ (red$^{33,75,80,81}$) as a function of redshift. The shaded regions shows estimates of the error on these values.  For   $\Omega_{\rm HI}$ the points above $z=0.24$ are determined from absorption line studies. Also shown is the the range 
of star formation rate density of galaxies as a function of redshift (green$^{82}$).}
\end{minipage}
\end{center}
\end{figure}

The required census of gas and stars needs to span a wide range of galaxy
environments, include a significant number of sources over all mass
intervals,
and ideally probe galaxies to at least $z=1$ where significant changes in
the star formation rate begin to occur. To date the stellar content of galaxies has been relatively well studied through large-scale redshift surveys$^{62,63}$,
particularly with SDSS$^{64,65}$.
By comparison, statistical measurements of gas in galaxies are
still in their infancy.

The best way to assess the reservoir of neutral gas associated with
galaxies is through blind 21-cm surveys. Significant progress has been made
since the first of these surveys; from the initial survey finding
37 galaxies$^{66}$, to the HIPASS survey which detected
5317 galaxies$^{67,68}$, to ongoing surveys with Arecibo which are on track to detect
over 25,000 galaxies$^{69,70}$.
Nevertheless, these surveys are low resolution and extend to only $z=0.06$, limiting their ability   to address the astrophysical questions that tie the cold atomic
gas to stars over cosmic time. Measuring the \HI\ content of galaxies beyond $z \sim$ 0.05 and making the
connection to the Ly-$\alpha$ absorbers is key for progress to be made$^{71,72}$.
The absorbers contain a significant amount of the baryons in the universe$^{73}$, but very little of this material is currently in neutral atomic gas.

The first 21-cm detections of massive galaxies at
$z\sim$0.2 have recently been made$^{74,75,76,77}$.
Observations in the next decade need to press beyond measuring
the gas content of a handful of the most massive systems and extend the gas census to increasingly higher redshifts where the census of stars has already been measured.  Combinations with molecular gas surveys and absorption line studies are also key to understand the link between a galaxy's future fuel, existing fuel, and stellar component.

\nsfsect{Key Advances}

In order for progress to be made in this field in the next decade key advances must be made in the observational realm, as well as parallel improvements in simulations.  

\smallskip
\noindent
{\bf Gas Accretion - }  How does a galaxy accrete gas?  What is the origin of the gas being accreted and how does reality compare to simulation results (e.g., Figure 1)?
An assessment of the number of galaxies with significant cold gaseous structures in their halos will make progress towards understanding the mode gas accretes onto galaxies.  This can be done with ongoing surveys with single-dish telescopes (e.g., Arecibo, GBT), but the resolution and sensitivity capabilities of these telescopes limit this work to the closest galaxies and often leave the origin of the gas unclear.  To detect and resolve typical Local Group halo clouds beyond a few Mpc, an expanded Allen Telescope Array (e.g., an ATA with 256 antennas$^{83}$) or expanded EVLA is needed$^{84}$.
Ultimately the Square Kilometer Array (SKA) will be required to detect halo clouds at distances beyond the Virgo Cluster and make the link to the lower column density gas traced in absorption$^{85}$.  The expanded ATA has the potential to reach \HI\ column densities of $10^{17}$ cm$^{-2}$, but it will take $\sim1000$ hrs of integration time on a single pointing. 
The SKA will also allow the morphology of the detected gas to be resolved at higher redshifts (i.e., coherent streams vs. halo merging).  The installation of the Cosmic Origins Spectrograph (COS) on HST will dramatically improve studies of diffuse gas in the universe through absorption line studies, but these studies
cannot distinguish between gas that is accreting versus outflowing, depict the environment of the gas, or determine the mass scale of the flow without corresponding emission line maps.   Sensitive high resolution \HI\ observations, combined with H$\alpha$ maps with wide-field IFU spectrographs, are essential to better understand the multi-phase accretion in galaxy halos and constrain models of galaxy evolution.  

The near future of the cosmological simulations is to model gas dynamics,
coupled with radiative transfer, with resolutions of tens of parsecs over the 
entire galaxy halo.  
Many questions remain about the physical effects that can influence the
coherent accretion flows and halo clouds, including:  galactic
outflows, the halo substructure, dynamical instabilities, magnetic
fields, conduction, local ionization fields, etc.
Some of these processes have been examined in high resolution, non-cosmological simulations,
but ultimately they need to be incorporated into fully cosmological
simulations where the hierarchical and complex nature of galaxy build-up can
be modeled in its full complexity.  

\smallskip
\noindent
{\bf Star Formation -} How does gas transition from atomic to molecular gas and subsequently form stars?  Multi-wavelength observations are required, combined with
theory and simulations, with both probing similar scales in a variety of environments.
%
Simulations need to include time-dependent chemistry and radiative
transfer to facilitate a more quantitative comparison to observations.
This will not only require new code development efforts, but also a
database including interfaces to access the simulation data.
The observations are currently limited to the closest galaxies and the most
intense star forming environments and need to expand the range of density regimes explored in order to better constrain the theory.  Optical, IR, and FIR imaging and spectroscopy
are necessary to characterize recent star formation in outer disks, determine the chemical enrichment history, and measure a galaxy's star formation rate.  A combination of sensitive single dish observations and interferometry in the radio, mm, and sub-mm regimes is needed to map the distribution and kinematics of the
\HI, trace the transition to molecular gas, and probe the role of dust over the full parameter space of
galaxy properties.  High resolution images of star formation tracers beyond the optical and UV would be a spectacular step forward in understanding the relationship between SFR and gas density. The telescopes needed over the next generation to continue to improve our understanding in this field include the EVLA, ALMA, CARMA, LMT, GBT, JWST, and SKA.

\smallskip
\noindent
{\bf Gas in the Universe -}  How does the census of gas in galaxies evolve with the census of stars?  What is the rapid refueling process implied by Figure 2?  Knowledge of the gas content of the universe has severely lagged our knowledge of the stellar component.   This lag has led to severe extrapolations based on limited data and hampers our ability to develop a coherent picture of galaxy evolution.
Ideally, future surveys will detect galaxies in \HI\ emission in clusters, filaments and voids out to the peak in the cosmic star formation rate density, i.e., the epoch of galaxy assembly at $z\sim2-3$.
A telescope with sufficient sensitivity to measure the \HI\ much beyond $z\sim1$ would have an order of magnitude or more collecting area than is currently
available.
The SKA, with a square kilometer of collecting area and state-of-the-art elements and design, would detect \HI\ emission from a Milky Way
to redshifts beyond 1.5, and with deep pointings to $z\sim3$.$^{85}$  
In the next few years surveys and telescopes can be completed
to provide a cosmic census of the distribution and kinematics of gaseous baryons out to $z\sim0.5$, 
tracing the build up of cosmic structure over the past $\sim4$ Gyr.
Arecibo's survey work of the local Universe can be extended to redshifts of a few tenths, and the 
EVLA at 21-cm can be used for deep integrations of narrow cones in redshift space from $z=0-0.5$$^{84}$,
with the full resolution of the VLA.
Building out the ATA to several hundred dishes would allow wide-field, fast surveys of \HI\ with several times
the depth of the Arecibo surveys, and better resolution$^{83}$. 
Through combinations with surveys of the molecular gas (using ALMA and LMT) and ongoing stellar surveys, substantial progress
can be made in the next decade on understanding how galaxies obtain gas and form stars in the evolving universe.

\bigskip
\noindent {\bf References:}
\footnotesize
$^1$Kere{\v s}, D., Katz, N., Weinberg, D.H. \& Dav{\'e}, R. 2005, MNRAS, 363, 2; $^2$Dekel, A. et al. 2009, Nature, 457, 451; $^3$Kere{\v s}, D., et al. 2009, MNRAS, in press (arXiv:0809.1430); $^4$Brooks et al., 2009, ApJ, in press (arXiv:0812.0007); $^5$White, S. \& Rees, M. 1978, MNRAS, 183, 341; $^6$Mo, H. \& Miralda-Escude, J. 1996, ApJ, 469, 589; $^7$Maller, A. \& Bullock, J. 2004, MNRAS, 355, 694; $^8$Kaufmann, T. et al. 2006, MNRAS, 370, 1612; $^9$Sommer-Larsen, J. 2006, ApJ, 644, L1; $^{10}$Wakker, B. \& van Woerden, H. 1997, ARA\&A, 35, 217; $^{11}$Putman, M. et al. 2002, AJ, 123, 873; $^{12}$Chiappini, C. Matteucci, F., \& Romano, D. 2001, ApJ, 554, 1044; $^{13}$Sancisi, R. 2008, A\&ARv, 15, 189; $^{14}$Thilker, D. et al. 2004, ApJ, 601, L39; $^{15}$Grcevich, J. \& Putman, M. 2009, ApJ, in press; $^{16}$McMahon, R. et al. 1990, ApJ, 359, 302; $^{17}$Fraternali, F., \& Binney, J. J. 2008, MNRAS, 386, 39; $^{18}$Peek, J., Putman, M. \& Sommer-Larsen, J. 2008, ApJ, 674, 227; $^{19}$Morganti, R. et al. 2006, MNRAS, 371, 157; $^{20}$van Gorkom, J. \& Schiminovich, D. 1997, ASP Conf. V. 116, 310;  $^{21}$Heitsch, F. \& Putman, M. 2009, ApJ, submitted;  $^{22}$Bland-Hawthorn, J. et al. 2008, ApJL; $^{23}$Sembach et al. 2003, ApJS, 146, 165; $^{24}$Rossa, J., \& Dettmar, R.-J. 2003, A\&A, 406, 493; $^{25}$Oosterloo, T., Fraternali, F., \& Sancisi, R. 2007, AJ, 134, 1019; $^{26}$Moore, B, et al. 1999, ApJ, 524, L19; $^{27}$Klypin, A., et al. 1999, ApJ, 522, 82; $^{28}$Bell, E. F., et al. 2005, ApJ, 625, 23; $^{29}$Somerville, R. S., et al. 2008, MNRAS, 391, 481; $^{30}$Kauffmann, G., et al. 2003, MNRAS, 341, 54; $^{31}$Schiminovich, D., et al. 2007, ApJS, 173, 315; $^{32}$Madau, P., et al. 1996, MNRAS, 283, 1388; $^{33}$Prochaska, J. X., Herbert-Fort, S., Wolfe, A. M. 2005, ApJ, 635, 123;  $^{34}$Madau, P., Ferrara, A., \& Rees, M. J. 2001, ApJ, 555, 92; $^{35}$Tremonti, C. et al. 2004, ApJ, 613, 898; $^{36}$Wolfe, A. M., \& Chen, H.-W. 2006, ApJ, 652, 981; $^{37}$Thilker, D. et al. 2007, ApJS, 173, 578; $^{38}$Meurer et al. 2009 ApJ, in press, arXiv:0902.0384; $^{39}$Schmidt, M. 1959, ApJ. 129. 243; 
$^{40}$Kennicutt, Jr., R. C. 1998, ApJ, 498, 541; $^{41}$Toomre, A. 1964, ApJ, 139, 1217; $^{42}$Li, Y., Mac Low, M.-M., \& Klessen, R. S. 2005, ApJ, 626, 823; $^{43}$Robertson, B. E., \& Kravtsov, A. V. 2008, ApJ, 680, 1083; $^{44}$Leroy, A. K., et al. 2008, AJ, 136, 2782; $^{45}$Dong, H., et al. 2008, AJ, 136, 479; $^{46}$Martin, C. L., \& Kennicutt, Jr., R. C. 2001, ApJ, 555, 301; 
$^{47}$Kim, W.-T., \& Ostriker, E. C. 2006, ApJ, 646, 213;  $^{48}$Hunter, D. A., Elmegreen, B. G., \& Baker, A. L. 1998, ApJ, 493, 595;  $^{49}$Tasker, E. J., \& Tan, J. C. 2008, ApJ, submitted (arXiv:0811.0207);  $^{50}$Wong, T., \& Blitz, L. 2002, ApJ, 569, 157; $^{51}$Heitsch, F., \& Hartmann, L. 2008, ApJ, 689, 290; $^{52}$Krumholz, M. R., \& McKee, C. F. 2005, ApJ, 630, 250; $^{53}$Mac Low, M. \& Klessen, R. 2004, RvMP, 76, 125; $^{54}$Boissier, S., et al. 2003, MNRAS, 346, 1215; $^{55}$Meurer, G. et al. 2006; ApJS, 165, 307; $^{56}$Kennicutt, Jr., R. C., et al. 2007, ApJ, 671, 333; $^{57}$Calzetti, D., et al. 2007, ApJ, 666, 870; $^{58}$Bigiel, F., et al. 2008, AJ, 136, 2846; $^{59}$Blitz, L. \& Rosolowsky, E. 2006, ApJ, 650, 933; $^{60}$Leroy, A. K., et al. 2007, ApJ. 658, 1027; $^{61}$Bolatto, A. D., et al. 2008, ApJ, 686, 948; 
$^{62}$Colless, M. et al. 2001, MNRAS, 328, 1039; $^{63}$Croton, D. et al. 2004, MNRAS, 356, 1155;  $^{64}$Blanton, M. R. et al. 2005, ApJ, 631, 208; $^{65}$Baldry, I. K. et al. 2005, MNRAS, 358, 441; $^{66}$Henning, P.A. 1992, ApJS, 78, 365; $^{67}$Meyer, M. J. et al. 2004, MNRAS, 350, 1195; $^{68}$Wong et al. 2006, MNRAS, 371, 1855; $^{69}$Giovanelli, R. et al. 2005, AJ, 130, 2598; $^{70}$Auld, R. et al. 2006, MNRAS, 371, 1617; $^{71}$Chen, H.-W. et al. 2005, ApJ, 629, L25; $^{72}$Ryan-Weber, E. 2006, MNRAS, 367, 1251; $^{73}$Danforth, C. W. \& Shull, J. M. 2008, ApJ, 679, 194; $^{74}$Verheijen, M. et al. 2007, ApJ, 668, L9; $^{75}$Lah, P. et al. 2007, MNRAS, 376, 1357; $^{76}$Catinella et al. 2008, ApJ, 685, 13; $^{77}$Zwaan, M. A., van Dokkum, P. G., Verheijen, M. A. W. 2001, Sci, 293, 1800; 
$^{78}$Drory et al. 2004, ApJ, 608, 742; $^{79}$Drory, N. et al. 2005, ApJ, 619, 131; $^{80}$Rao, S. M., Turnshek, D. A., Nestor, D. B. 2006, ApJ, 636, 610; $^{81}$Zwaan, M. et al., 2005, MNRAS, 359, L30; $^{82}$Hopkins, A. et al. 2004, ApJ, 615, 209; $^{83}$etoile.berkeley.edu$/$$\sim$gbower$/$RSS$/$rss\_08dec15.pdf; $^{84}$www.aoc.nrao.edu$/$evla; $^{85}$www.astro.cornell.edu$/$ska; www.skatelescope.org.

\end{document}